\documentclass[twocolumn,aps,pra,reprint,superscriptaddress,longbibliography]{revtex4-1}
\usepackage{amsmath}
\usepackage{graphicx}
\usepackage[lofdepth,lotdepth, caption=false]{subfig}
\usepackage{verbatim}
\usepackage{color}
\usepackage{dcolumn}
\usepackage{bm}
\usepackage{miller}
\usepackage{color}

\usepackage{xr-hyper}
\usepackage{hyperref}

\usepackage[english]{babel}

\begin{document}

\title{Large change of interlayer vibrational coupling with stacking in Mo$_{1-x}$W$_{x}$Te$_{2}$}

\author{John A.~Schneeloch}
\affiliation{Department of Physics, University of Virginia, Charlottesville,
Virginia 22904, USA}

\author{Yu Tao}
\affiliation{Department of Physics, University of Virginia, Charlottesville,
Virginia 22904, USA}

\author{Jaime A.\ Fernandez-Baca}
\thanks{Notice:  This manuscript has been authored by UT-Battelle, LLC, under contract DE-AC05-00OR22725 with the US Department of Energy (DOE). The US government retains and the publisher, by accepting the article for publication, acknowledges that the US government retains a nonexclusive, paid-up, irrevocable, worldwide license to publish or reproduce the published form of this manuscript, or allow others to do so, for US government purposes. DOE will provide public access to these results of federally sponsored research in accordance with the DOE Public Access Plan (http://energy.gov/downloads/doe-public-access-plan).}
\affiliation{Neutron Scattering Division, Oak Ridge National Laboratory, Oak Ridge, Tennessee 37831, USA}

\author{Guangyong Xu}
\affiliation{NIST Center for Neutron Research, National Institute of Standards and Technology, Gaithersburg, Maryland 20877, USA}

\author{Despina Louca}
\thanks{Corresponding author}
\email{louca@virginia.edu}
\affiliation{Department of Physics, University of Virginia, Charlottesville,
Virginia 22904, USA}

\begin{abstract}
Stacking variations in quasi-two-dimensional materials can have an important influence on material properties, such as changing the topology of the band structure. 
Unfortunately, the weakness of van der Waals interactions makes it difficult to compute the stacking dependence of properties, and even in a material as simple as graphite the stacking energetics remain unclear.
Mo$_{1-x}$W$_{x}$Te$_{2}$ is a material in which three differently-stacked phases are conveniently accessible by temperature changes: $1T^{\prime}$, $T_d^*$, and the reported Weyl semimetal phase $T_d$.
The transitions proceed via layer sliding, and the corresponding interlayer shear mode (ISM) is relevant not just for the stacking energetics, but for understanding the relationship between the Weyl physics and structural changes. 
However, the interlayer interactions of Mo$_{1-x}$W$_{x}$Te$_{2}$ are not well understood, with wide variation in computed properties. 
We report inelastic neutron scattering of the ISM in a  Mo$_{0.91}$W$_{0.09}$Te$_{2}$ crystal.
The ISM energies are generally consistent with the linear chain model (LCM), as expected given the weak interlayer interaction, though there are some discrepancies from predicted intensities.
However, the interlayer force constants $K_x$ in the $T_d^*$ and $1T^{\prime}$ phases are substantially weaker than that of $T_d$, at 76(3)\% and 83(3)\%, respectively.
Considering that the relative positioning of atoms in neighboring layers is approximately the same regardless of overall stacking, our results suggest that longer-range influences, such as stacking-induced electronic band structure changes, may be responsible for the substantial change in the interlayer vibrational coupling and, thus, the $C_{55}$ elastic constant. 
These findings should elucidate the stacking energetics of Mo$_{1-x}$W$_{x}$Te$_{2}$ and other van der Waals layered materials.

\end{abstract}

\maketitle

\section{Introduction}

Variations in the layer stacking of quasi-two-dimensional (quasi-2D) materials can sometimes have important effects on material properties. For example, the chromium trihalides CrX$_{3}$ (X=Cl, Br, I) have interlayer magnetic coupling that changes with layer stacking \cite{klein_enhancement_2019,chen_direct_2019,li_pressure-controlled_2019}, and MoTe$_{2}$ is reported to be a Weyl semimetal in its low-temperature $T_{d}$ phase but not in its higher-temperature $1T^{\prime}$ phase \cite{sun_prediction_2015,deng_experimental_2016}. These materials are also examples where stacking changes can be conveniently induced by modifying an external parameter such as temperature \cite{clarke_low-temperature_1978,mcguire_crystal_2017}. 
Unfortunately, theoretical investigation of these transitions and the stacking dependence of properties is hindered by the weakness of the interlayer van der Waals (vdW) interactions, which results in small energy differences between stacking variations and increases the precision needed for calculations. 
Even in a material as simple and as frequently studied as graphite, there have been scant experimental and contradictory theoretical studies on whether the rhombohedral or Bernal stacking has a lower free energy at room temperature \cite{nery_ab-initio_2021}. 
Experiments where properties are measured across stacking variations could provide much needed insight into interlayer interactions and stacking energetics. 

In MoTe$_{2}$, one can switch between three different layer stacking orders by changes in temperature \cite{tao_appearance_2019,schneeloch_evolution_2020}. MoTe$_{2}$ crystallizes in the monoclinic $1T^{\prime}$ phase, which can be preserved at room temperature over the more stable 2H phase by quenching \cite{clarke_low-temperature_1978}. On cooling $1T^{\prime}$ below $\sim$280 K,  disordered stacking appears, with a gradual transition into the orthorhombic $T_{d}$ phase. On warming above $\sim$260 K, $T_{d}$ abruptly transitions into the pseudo-orthorhombic $T_{d}^{*}$ phase, and further warming results in disordered stacking with a gradual transition back into the $1T^{\prime}$ phase. W substitution up to $x \sim 0.2$ results in increased transition temperatures but similar transitions \cite{schneeloch_evolution_2020}.

The interlayer interaction between neighboring layers can be thought of as a double-well potential \cite{heikes_mechanical_2018}, where the minima correspond to two stacking options, which we label ``A'' and ``B'', that are accessible by layer sliding along the $a$-axis (Fig.\ \ref{fig:1}(a,b)).  
The multitude of stacking configurations accessible from $1T^{\prime}$ via temperature changes are all constructible by an A/B sequence of stacking operations \cite{schneeloch_evolution_2020,schneeloch_emergence_2019}. 
For example, repeated AA...\ stacking yields $T_{d}$, AABB...\ yields $T_{d}^{*}$, and AB...\ yields $1T^{\prime}$.
Performing an inversion operation reverses the A/B stacking sequence while swapping every ``A'' with ``B'' and vice versa; for example, inversion of the $T_{d}$ twin with AA...\ stacking results in the other $T_{d}$ twin, which has BB...\ stacking. 
Thus, for $T_{d}$, the A and B stacking operations are symmetry-equivalent, and this statement can be extended to all A/B stacking sequences under the assumption of identical and centrosymmetric layers \cite{schneeloch_evolution_2020}. (This assumption is justified by the fact that differences in the intralayer positioning of atoms between, e.g.,  $1T^{\prime}$-MoTe$_{2}$ and $T_d$-MoTe$_{2}$ are $\lesssim$ 0.5\% of the lattice constants, as seen from reported coordinates in, e.g., Ref.\ \cite{heikes_mechanical_2018}.) Thus, to a first approximation, we should expect interlayer vibrational coupling between neighboring layers to be similar regardless of overall stacking. 

The $a$-axis interlayer shear mode (ISM) has been studied for its relevance in identifying the $T_d$ phase and in modulating its Weyl semimetal properties.
These studies include Raman spectroscopy in MoTe$_{2}$ \cite{zhang_raman_2016, ma_raman_2016, chen_activation_2016, cao_barkhausen_2018, cheon_structural_2021} and WTe$_{2}$ \cite{xiao_berry_2020, kim_determination_2016, jiang_raman_2016}, and various ultrafast spectroscopy techniques in MoTe$_{2}$ \cite{zhang_light-induced_2019,fukuda_ultrafast_2020,qi_photoinduced_2021,rivas_generation_2019} and WTe$_{2}$ \cite{he_coherent_2016,soranzio_ultrafast_2019,hein_mode-resolved_2020,drueke_observation_2021,qi_photoinduced_2021,ji_manipulation_2021,sie_ultrafast_2019}. 
Raman spectroscopy, however, is limited to measuring the zone-center energy $\hbar \omega_m$ (i.e., the maximum of the ISM dispersion), and only in the $T_d$ phase (for bulk samples) is this mode Raman-active. 
The ultrafast spectroscopy techniques involve firing a femtosecond light pulse at the sample, then measuring the picosecond-scale changes in the intensity of electron diffraction, reflectivity, second harmonic generation, angle-resolved photoemission spectroscopy (ARPES), etc., frequently in the form of oscillations of angular frequency $\omega_m$. 
These techniques have provided much insight into the connection between the electronic topology and the structure; for instance, modulations in electronic states near the Weyl node locations with the oscillation of the interlayer shear mode in MoTe$_{2}$ have been observed via ARPES \cite{hein_mode-resolved_2020}, and a link between the Weyl fermions and relaxation dynamics of this mode has been suggested \cite{drueke_observation_2021}. 
However, ultrafast spectroscopy techniques may have complications such as the fluence- and pump-frequency-dependence of observed mode frequencies \cite{sie_ultrafast_2019}.

Meanwhile, theoretical studies on MoTe$_{2}$ and WTe$_{2}$ have had wide discrepancies on properties relevant to the interlayer interactions, such as values of $\omega_{m}$ or the $a$-axis displacement between the A/B stacking options. Experimentally, $\hbar \omega_{m}$ for MoTe$_{2}$ has been reported from Raman spectroscopy as 1.61 meV (10 K, $T_d$) \cite{ma_raman_2016} or 1.56 meV (78 K, $T_d$) \cite{chen_activation_2016}, and from ultrafast spectroscopy as 1.61 meV (300 K, $1T^{\prime}$) \cite{fukuda_ultrafast_2020} and 1.74 meV ($\leq$ 240 K, $T_d$) \cite{zhang_light-induced_2019}. Density functional theory (DFT) calculations, on the other hand, have resulted in much wider variation, with values of 1.40 \cite{heikes_mechanical_2018}, 1.28 \cite{ma_raman_2016}, and 1.14 meV \cite{chen_activation_2016}  for $T_d$-MoTe$_{2}$, and 1.09 \cite{ma_raman_2016} and 1.90 meV \cite{chen_activation_2016} for $1T^{\prime}$-MoTe$_{2}$. %
The elastic constant $C_{55}$ describes the resistance to shear strain in the long-wavelength limit of the ISM. For $T_d$-MoTe$_{2}$, $C_{55}$ has been calculated as 24.3 \cite{rano_ab_2020} and 3.9 GPa \cite{singh_engineering_2020}, and for $1T^{\prime}$-MoTe$_{2}$ as 2.9 GPa \cite{singh_engineering_2020}, which imply (via the linear chain model, to be discussed below) $\hbar \omega_m$ values of 3.34, 1.34, and 1.15 meV, respectively. 
The layer-sliding distance $\epsilon$ between the A/B stacking options
also tends to be underestimated in DFT calculations (e.g., the calculated $\beta$ angles of $1T^{\prime}$-MoTe$_{2}$ and $1T^{\prime}$-WTe$_{2}$ in Ref.\ \cite{kim_origins_2017} are lower than the experimental values \cite{clarke_low-temperature_1978,tao_t_d_2020}.)
Inelastic neutron scattering (INS) is uniquely useful as a probe of phonons across a range of momentum transfers, and can yield insights on the interlayer phonons of Mo$_{1-x}$W$_{x}$Te$_{2}$ beyond that estimated via DFT or reported in Raman or ultrafast spectroscopy measurements.

We present inelastic neutron scattering  measurements on a Mo$_{0.91}$W$_{0.09}$Te$_{2}$ crystal, measuring the ISM mode in the $T_d$, $T_d^*$, and $1T^{\prime}$ phases. 
The phonon energies are consistent with a linear chain model (LCM), but the interlayer force constants for $T_d^*$ and $1T^{\prime}$ are, respectively, about 76(3)\% and 83(3)\% that of the $T_d$ phase. 
The large change in the force constants for different stacking orders, in contrast to the minimal change in the relative positioning of neighboring layers regardless of stacking, suggests that stacking-induced electronic band structure changes may play a substantial role in the interlayer vibrational coupling.

\section{Experimental Details}

Inelastic neutron scattering was performed on a $\sim$0.6 g Mo$_{0.91}$W$_{0.09}$Te$_{2}$ crystal, labeled ``MWT1'' and measured in previous neutron scattering studies \cite{tao_appearance_2019, schneeloch_evolution_2020}. The W fraction in Mo$_{1-x}$W$_{x}$Te$_{2}$ was estimated to be $x \approx 0.09(1)$ from the interlayer spacing obtained from the position of the $(004)$ peak in neutron scattering measurements, roughly consistent with the $x \approx 0.06(1)$ value obtained via energy-dispersive x-ray spectroscopy measurements of the surface. A second $\sim$0.1 g crystal, labeled MT2 and having composition Mo$_{1-x}$W$_{x}$Te$_{2}$ with $x \leq 0.01$ \cite{tao_appearance_2019,schneeloch_evolution_2020}, was used for a single measurement. MWT1 and MT2 were grown from a Te flux; details can be found in Ref.\ \cite{schneeloch_evolution_2020,tao_appearance_2019}.

Cold-neutron triple axis spectrometer measurements were performed at the CTAX instrument at the High Flux Isotope Reactor of Oak Ridge National Laboratory, and on the SPINS instrument at the NIST Center for Neutron Research at the National Institute of Standards and Technology. Final neutron energy was fixed at 4.5 and 5.0 meV for CTAX and SPINS, respectively. The collimations were 48$^{\prime}$-40$^{\prime}$-S-40$^{\prime}$-120$^{\prime}$ for CTAX and open-80$^{\prime}$-S-80$^{\prime}$-open for SPINS. For CTAX, a Be filter was used after the sample. For SPINS, Be filters were used before and after the sample. For all analyzer and monochromator crystals, the $(002)$ plane of pyrolytic graphite was used. 

For simplicity, we present all data in the  $T_{d}$-phase reciprocal space coordinates based on an orthorhombic unit cell with $a \approx 6.3$ \AA, $b \approx 3.47$ \AA, and $c \approx 13.8$ \AA, regardless of the phase being measured. The intensities for the data from a particular instrument share the same arbitrary units. Error bars denote a standard deviation of statistical uncertainty.

\section{Results}

\begin{figure}[h]
\begin{center}
\includegraphics[width=8.6cm]
{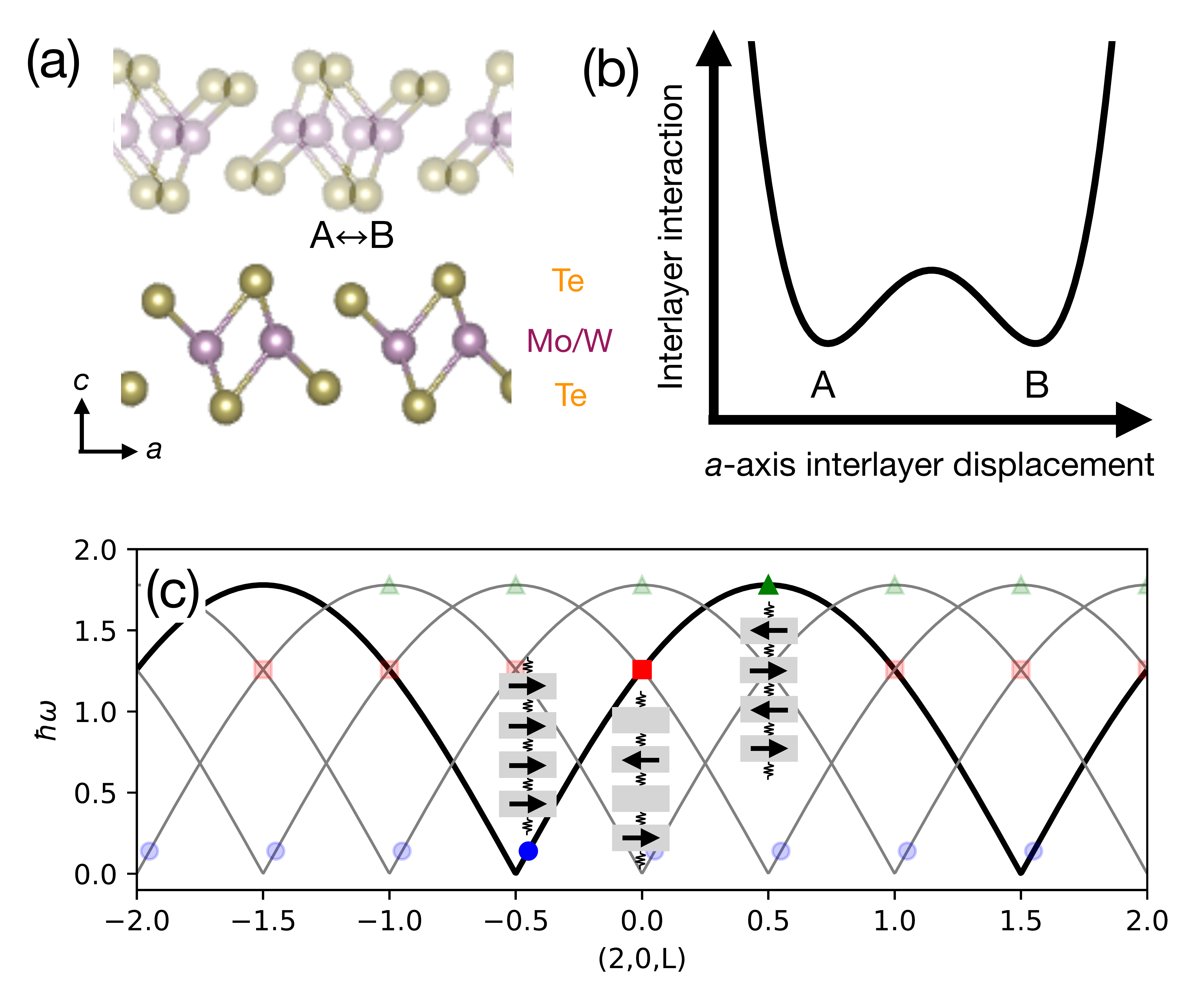}
\end{center}
\caption{(a) Crystal structure of Mo$_{1-x}$W$_{x}$Te$_{2}$, with A/B stacking options displayed. (b) Diagram of interlayer interaction energy as a function of relative displacement of neighboring layers along the $a$-axis. 
(c) A depiction of the dispersion along $(2,0,L)$ for the $a$-axis interlayer shear mode based on the linear chain model for a four-layer unit cell (i.e., $T_d^*$). One sub-branch of the LCM dispersion is made bold. The sets of blue circles, red squares, and green triangles each mark a particular vibrational mode on the LCM curve, and are accompanied by diagrams of the polarization of the interlayer vibrations, depicting the relative phases $(...,1,1,1,1,...)$, $(...,1,-i,-1,i,...)$, and $(...,1,-1,1,-1,...)$, respectively.
}
\label{fig:1}
\end{figure}

\begin{figure}[h]
\begin{center}
\includegraphics[width=8.6cm]
{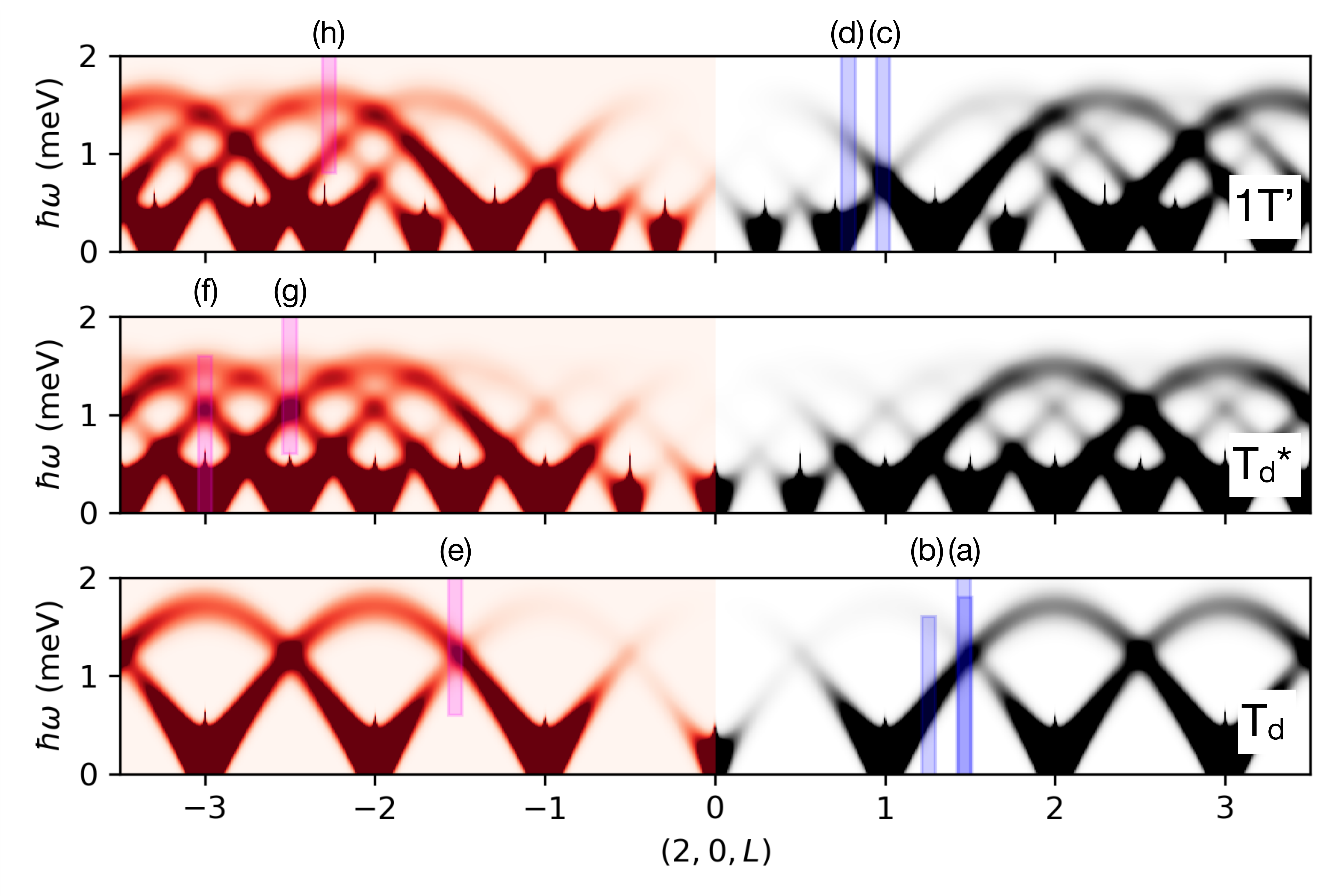}
\end{center}
\caption{
Calculated inelastic neutron scattering intensity for each phase as determined by the LCM, setting $T= 270$ K and $\hbar \omega_{m}$ to the values for each phase listed in Table \ref{tab:freqAve}. Intensity convoluted with an energy FWHM of 0.3 meV. The left (right) shows the intensity for the $T_d^*$/$1T^{\prime}$ twin fractions derived from elastic scans taken on the SPINS (CTAX) instrument. The blue and pink bars denote scans taken on CTAX and SPINS, respectively. The letters refer to the data sets in Fig.\ \ref{fig:3}.
}
\label{fig:2}
\end{figure}

\begin{figure}[h]
\begin{center}
\includegraphics[width=8.6cm]
{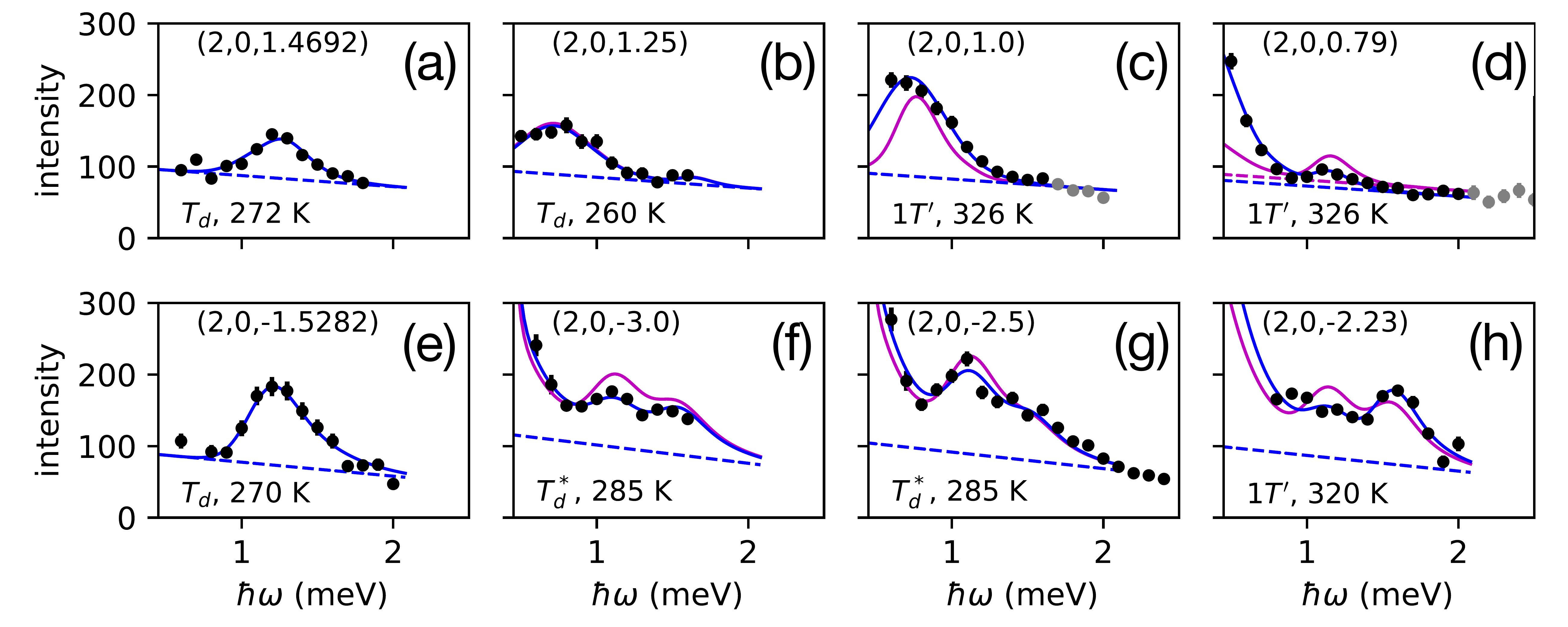}
\end{center}
\caption{
(a-h) Scans of inelastic neutron scattering intensity vs.\ $\hbar \omega$ taken on (a-d) CTAX and (e-h) SPINS, as labeled in Fig.\ \ref{fig:2}. 
Blue and magenta curves are resolution-convoluted $S(\mathbf{Q},\omega)$ calculations. For the blue curves, intensity, twin fraction, and $\hbar \omega_m$ were allowed to vary. For magenta curves, intensities in (b-d) and (f-h) were constrained by the LCM and fitted intensities of (a) and (e); twin fractions were set to values consistent with elastic $(2,0,L)$ scans; and $\hbar \omega_m$ was set to the average values for each phase listed in Table \ref{tab:freqAve}. Dashed lines are background. Gray points are data not included in fit.
}
\label{fig:3}
\end{figure}

\begin{figure}[h]
\begin{center}
\includegraphics[width=8.6cm]
{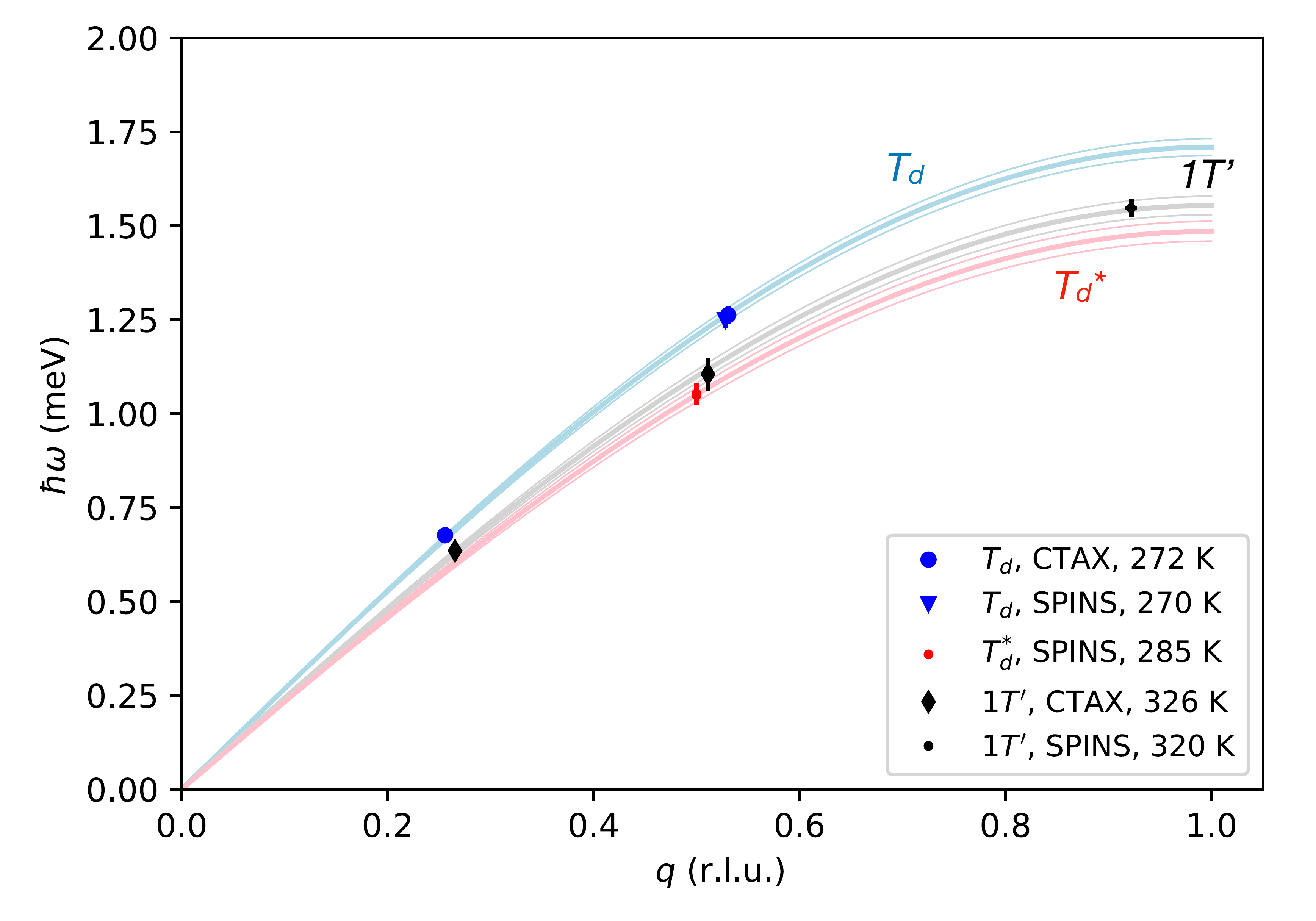}
\end{center}
\caption{Comparison of linear chain model with Mo$_{0.91}$W$_{0.09}$Te$_{2}$ neutron scattering data. The data points are  $\hbar \omega_m \sin(\frac{\pi}{2} q)$ plotted against $q$, where $q$ is the LCM wavevector corresponding to the branch that dominates the contribution to the intensity, and $\hbar \omega_m$ are the values obtained from fits which are shown in Table \ref{tab:maxFreq}.
The LCM curves are $\hbar \omega = \hbar \omega_{m} \sin{\frac{\pi}{2} q}$ for each phase, with $\hbar \omega_m$ given by the values in Table \ref{tab:freqAve}. 
The side curves show changes in the LCM curve by a standard deviation in $\hbar \omega_m$.
}
\label{fig:4}
\end{figure}

The linear chain model is often used in studying interlayer vibrational modes of quasi-2D materials, especially in the context of Raman spectroscopy measurements on few-layer crystals \cite{liang_low-frequency_2017}. The LCM  represents interlayer vibrational coupling as if the layers were particles coupled by springs to their neighbors. 
For an infinite chain, the dispersion is given by 
\begin{equation}
    \label{eq:LCM}
    \hbar \omega_q = 2 \hbar \sqrt{ \frac{K_x}{\mu}} \left|\sin{\frac{\pi q}{2}}\right|,
\end{equation}
where $q$ is the LCM wavevector (scaled such that $q=1$ at the BZ boundary, with $q$ in the same r.l.u.\ as $L$), $\hbar \omega_q$ is the phonon energy, $K_x$ is the interlayer force constant for the ISM, $\mu$ is the areal mass density per layer, and $\hbar$ is Planck's constant divided by $2 \pi$. The only free parameter in this model is the ratio $K_x/\mu$. 

The LCM dispersion measured by neutron scattering has complications over the $\left|\sin \frac{\pi q}{2}\right|$ form due to layers having differing orientation and in-plane positioning. To illustrate, Fig.\ \ref{fig:1} depicts the dispersion for the four-layer unit cell of $T_d^*$ along $(2,0,L)$, in which the $\left|\sin \frac{\pi q}{2}\right|$ dispersion is ``folded back'' every half-integer $L$, resulting in four different sub-branches repeated every half-integer $L$. (This dispersion can also be interpreted as joined acoustic/optic branches.)
To compute the expected phonon intensity for $T_d^*$, we employ our core LCM assumption, which is that the polarization vectors are uniform within each layer, are aligned exclusively along the $a$-axis, and have the LCM phases $\frac{1}{\sqrt{N}} e^{-i \pi l q} = \frac{1}{\sqrt{N}} e^{-i \pi l (L-L_0)}$, where $l = 0,...,N-1$ is the layer index, $N$ is the number of layers in the unit cell, and $L_0$ is a multiple of 1/2 corresponding to a $T_d^*$ Bragg peak location $(2,0,L_0)$. 
The integrated intensity of a phonon peak in a constant-$\mathbf{Q}$ scan for $\hbar \omega > 0$ at temperature $T$ is proportional to $\frac{1}{\omega} |F(\mathbf{Q})|^2 (n(\omega,T) + 1)$ \cite{shirane_neutron_2002}. The quantity $n(\omega, T)$ is the Bose factor, and $F(\mathbf{Q})$ is the dynamic structure factor, given by
\begin{equation}
F(\mathbf{Q}) = \sum_j \frac{b_j}{\sqrt{m_j}} (\mathbf{Q} \cdot \mathbf{\xi}^s_j) e^{i \mathbf{Q} \cdot \mathbf{d_j}}.
\end{equation}
The index $j$ runs over each atom in the unit cell; $b_j$ are the nuclear scattering lengths; $m_j$ and $\mathbf{d}_j$ are the masses and positions for atom $j$; $s$ labels a sub-branch; and $\mathbf{\xi}^s_j$ are the phonon polarization vectors.
(We neglect the Debye-Waller factor, which is $\sim$1 in the region of interest.) 
The expected LCM-derived INS intensity for the $T_d$, $T_d^*$, and $1T^{\prime}$ phases is shown in Fig.\ \ref{fig:2}. 
The $T_d$ and $1T^{\prime}$ phases fold back every integer $L$ away from their Bragg peaks due to their two-layer unit cells, but $1T^{\prime}$ has the additional complication that the intensity for each twin is shifted along $L$ by $\pm 2 \epsilon$ due to its monoclinic symmetry, with $\epsilon$ ($\sim$ 0.147 at 320 K \cite{schneeloch_evolution_2020})  being the $a$-axis displacement  between the two stacking options. 
The $T_d^*$ phase also has differing INS intensity for each twin, though the dispersion overlaps since the structure is pseudo-orthorhombic. For $T_d$, meanwhile, both twins produce identical INS intensity.

\begin{table}[t]
\caption{Values of $\hbar \omega_{m}$ obtained from fitting. ``Label'' corresponds to one of the data sets in Fig.\ \ref{fig:3}, except for ``MT2'' which denotes the data set corresponding to the MT2 sample. Nominal coordinates, phase, temperature, and the instrument used are also tabulated.}
\label{tab:maxFreq}
\begin{ruledtabular}
\begin{tabular}{llllll}
label & coordinates & phase & T (K) & inst. &  $\hbar \omega_{m}$ (meV)  \\
\hline
(a)    & (2,0,1.47) & $T_d$ & 272 & CTAX & 1.71(3)\\
  &  (2,0,1.46) & $T_d$ & 194 & CTAX & 1.76(6) \\
MT2  & (2,0,1.49) & $T_d$ & 232 & CTAX & 1.77(9) \\
(b)    & (2,0,1.25) & $T_d$ & 260 & CTAX & 1.74(5)\\
(e)    & (2,0,-1.53) & $T_d$ & 270 & SPINS & 1.694(29)\\
\hline
(f)    & (2,0,-3.0) & $T_d^*$ & 285 & SPINS & 1.49(4)\\
(g)    & (2,0,-2.5) & $T_d^*$ & 285 & SPINS & 1.48(6)\\
\hline
(c)    & (2,0,1.0) & $1T^{\prime}$ & 326 & CTAX & 1.57(3) \\
(d)    & (2,0,0.79) & $1T^{\prime}$ & 326 & CTAX & 1.512(20)\\
(h)    & (2,0,-2.23) & $1T^{\prime}$ & 320 & SPINS & 1.55(3)\\
  & (2,0,-2.23) & $1T^{\prime}$ & 500 & SPINS & 1.472(21) \\
  & (2,0,-2.23) & $1T^{\prime}$ & 600 & SPINS & 1.422(14) \\
\end{tabular}
\end{ruledtabular}
\end{table}

\begin{table}[t]
\caption{Values of $\hbar \omega_m$ for each phase, obtained from averaging within each phase the values of $\hbar \omega_m$ listed in Table \ref{tab:maxFreq}. Interlayer force constants $K_x$ and the ratios $K_x / K_x^{T_d}$ are also included.}
\label{tab:freqAve}
\begin{ruledtabular}
\begin{tabular}{llllll}
phase & $\hbar \omega_m$ (meV) & $K_x$ (10$^{19}$ N/m$^{3}$) & $K_x / K^{T_d}_x$ \\
\hline
$T_d$         & 1.709(22) & 0.919(24)  &   \\
$T_d^*$       & 1.486(26) & 0.694(25)      & 76(3)\% \\
$1T^{\prime}$ & 1.554(25) & 0.760(24)  & 83(3)\% \\ 
\end{tabular}
\end{ruledtabular}
\end{table}

We conducted scans of neutron scattering intensity along energy transfer $\hbar \omega$ at various points $L$ along $(2,0,L)$, as shown in Fig.\ \ref{fig:3}. (A few additional scans at different temperatures and on the MT2 crystal are shown in the Supplemental Materials \cite{supplement}.) Elastic scans along $(2,0,L)$ \cite{supplement} were taken before or after the inelastic scans to account for errors due to thermal expansion or changes in alignment. The curves in Fig.\ \ref{fig:3} show calculated $S(\mathbf{Q},\omega)$ convoluted with the instrument resolution. The blue curves are the result of fits where the overall intensity, $1T^{\prime}$/$T_d^*$ twin fraction, and dispersion maximum $\hbar \omega_m$ were allowed to vary, except for (c), in which the twin fraction was set to 100\% of the BA-stacked $1T^{\prime}$ twin. There is no obvious sign of broadening beyond the instrument resolution. 
(In these calculations, we relied on computed elastic constants \cite{rano_ab_2020} to estimate the dispersion of the ISM perpendicular to the $(2,0,L)$ line. We also estimated the in-plane sample mosaic from an analysis of our elastic $(2,0,L)$ scans. Inaccuracies in these assumptions could introduce systematic errors in the fitted $\hbar \omega_m$ values, though the ratios of $\hbar \omega_m$ between the phases is largely unchanged. See the Supplemental Materials for these and other fitting details, as well as why the layer breathing longitudinal acoustic mode can be neglected \cite{supplement}.)

The fitted $\hbar \omega_m$ values are shown in Table \ref{tab:maxFreq}, and show remarkable consistency within each phase. 
This consistency can be better seen in the plot of $\hbar \omega_m \sin(\frac{\pi q}{2})$ vs.\ $q$ in Fig.\ \ref{fig:4}, where $q$ is the LCM wavevector from the LCM sub-branch with the dominant contribution to the intensity. 
The two $T_d$ points near $q=0.53$ (corresponding to data sets (a) and (e)) overlap, and the point near $q=0.25$ (from (b)) is also consistent with the LCM curve. The two $T_d^*$ scans result in overlapping points near $q=0.5$, and the three $1T^{\prime}$ points are all consistent with the same curve.  
Averages within each phase of the fitted $\hbar \omega_m$ values are shown in Table \ref{tab:freqAve}, with $\hbar \omega_m = 1.709(22)$, 1.486(26), and 1.554(25) meV for the $T_d$, $T_d^*$, and $1T^{\prime}$ phases, respectively. 
The nearly-undoped crystal MT2 in its $T_d$ phase has a value of $\hbar \omega_m =  1.77(9)$ meV, consistent with $T_d$-MWT1. (The W-fraction dependence of $\hbar \omega_m$ can be estimated assuming a linear relation from reported values on MoTe$_{2}$ and WTe$_{2}$ \cite{ma_raman_2016}, yielding an expected decrease of $\sim$0.06 meV from MoTe$_{2}$ to MWT1, consistent with observations.)
The interlayer force constants $K_x$ are also listed, and we see that the $T_d^*$ and $1T^{\prime}$ phases have values of $K_x$ which are, respectively, $\sim$76\% and 83\% that of $T_d$. 
Thus, the vibrational coupling of the ISM is substantially weaker in $T_d^*$ and $1T^{\prime}$ than in $T_d$. This is remarkable considering that the $a$-axis displacement $\epsilon$ between the two stacking options is almost unchanged between $T_d$ and $1T^{\prime}$ (as can be seen from the discussion of the parameter $\delta=(\epsilon+1)/2$ parameter in Ref.\ \cite{schneeloch_evolution_2020}.)

Some temperature-induced phonon softening can be seen in our data, but the rate is far too low to account for the changes in $\hbar \omega_m$ between the phases. 
From the decrease in $\hbar \omega_m$ on warming from 320 to 600 K (in the $1T^{\prime}$ phase) for data taken near $(2,0,-2.23)$, we can estimate the softening rate to be -3.3(7)$\cdot$10$^{-4}$ K$^{-1}$.
Softening of the interlayer phonons is expected due to the known anharmonicity of the interlayer interaction \cite{heikes_mechanical_2018}, and would be consistent with the gradual reduction with warming in the spacing between the local minima (i.e., in $\epsilon$) \cite{schneeloch_evolution_2020}. 
Softening of the ISM modes has also been observed in WTe$_{2}$, where the relative change in $\omega_{m}$ per Kelvin is roughly $-4 \cdot 10^{-4}$ K$^{-1}$ within the range $0 \leq T \leq 300$ K \cite{he_coherent_2016}, a magnitude comparable to that in our data on $1T^{\prime}$-Mo$_{0.91}$W$_{0.09}$Te$_{2}$. 
Interestingly, a substantially greater softening was seen for the layer-breathing longitudinal acoustic mode in thin film MoTe$_{2}$, at $-2.0(1) \cdot 10^{-3}$ K$^{-1}$ \cite{rivas_generation_2019}.
In any case, a softening rate of the magnitude seen from 320 to 600 K is insufficient to explain the energy difference in the phonons between the $T_d$ and $T_d^*$ phases. If the rate were, say, $-4 \cdot 10^{-4}$ K$^{-1}$, we would only expect $\hbar \omega_{m}$ to decrease by about $-0.006$ meV from 270 to 285 K, or $-0.02$ meV from 270 to 320 K, far less than the 0.16 and 0.22 meV differences seen between $T_d$ and the other two phases. Thus, it is clear that the large changes in the interlayer force constant are due to changes in stacking. %
Such abrupt changes in $\hbar \omega_m$ can also be seen in Raman spectroscopy data on 22 and 155 nm thick MoTe$_{2}$ crystals
\cite{cao_barkhausen_2018}. (We note that, for the $T_d^*$ phase, \emph{two} interlayer force constants are allowed by symmetry in the LCM, but we expect little difference in intensity from a single-spring-constant model with an average value $K_x = \sqrt{K^1_x K^2_x}$, even if $K_x^1$ and $K_x^2$ differed by $\sim$20\%; see Supplemental Materials for details \cite{supplement}.)

While the energies are largely consistent with the LCM, the LCM-calculated intensities are not fully consistent with the data, which is especially evident for the $1T^{\prime}$ phase. The magenta curves in Fig.\ \ref{fig:3} are the ideal LCM $S(\mathbf{Q},\omega)$, in which the intensity for each instrument was set to the value determined from the $T_d$ measurements in sets (a) and (e), and kept fixed for the remaining data sets in (b-d) and (f-h). The twin fractions were set according to an analysis of elastic $(2,0,L)$ scan intensity \cite{supplement}, and $\hbar \omega_m$ was set for each phase to the values listed in Table \ref{tab:freqAve}. 
For the $T_d^*$ data, predicted intensities in the fitting ranges are somewhat greater than observed, though changes in the sample mosaic between phases could plausibly explain an overall decrease in intensity. There is a significant difference between the effective $T_d^*$ twin fractions needed to reproduce the (f) and (g) data (54(5)\% and 84(8)\% of the AABB twin, respectively), and the 70\% fraction that is consistent with the elastic data. 
However, for $1T^{\prime}$, the effective twin fractions needed to reproduce the inelastic intensity  ($\sim$100\%, 70(6)\%, and 81(3)\% of the BA twin for data sets (c), (d), and (h), respectively) are much different from the twin fractions consistent with the elastic data (25\%, 25\%, and 65\%), suggesting a substantial deviation from the linear chain model.
Such a deviation may be especially clear in $1T^{\prime}$ due to the twins of that phase having distinct peaks in much of the inelastic data, as opposed to the overlapping intensities of the $T_d^*$ twins. 
Regardless, such a large deviation suggests that the polarization vectors deviate from our assumption of uniformity within each layer, with a significant degree of intralayer vibrational motion, even if the mode energies remain consistent with the linear chain model.

\section{Discussion}
In a way, the structure of Mo$_{1-x}$W$_{x}$Te$_{2}$ is simple: Given identical, centrosymmetric layers, the layers are stacked according to an A/B sequence which determines whether the inversion symmetry centers of each layer are displaced by $+\delta$ or $-\delta$ along the $a$-axis relative to those of the layer below.
Differences in intralayer positioning between layers are small ($\leq$0.5\% of the lattice constants, as mentioned in the Introduction), and the parameter $\delta = (\epsilon - 1)/2$ appears to be practically unchanged between $T_d$ and $1T^{\prime}$ after accounting for an overall trend of a decrease in $\delta$ (or $\epsilon$) on warming  \cite{schneeloch_evolution_2020}. 

Nevertheless, our results indicate a large ($\sim$20\%) change in the interlayer shear vibrational coupling $K_x$ between $T_d$ and the other two phases. 
Presumably, while steric short-range interactions determine the relative $a$-axis displacement of the layers, the vibrational coupling depends strongly on the overall stacking of the layers, possibly through changes in the electronic band structure. 
(There may be an interesting correlation between the interlayer vibrational coupling and the resistivity. The (in-plane) resistivity appears to jump during the $T_d$$\rightarrow$$T_d^*$ transition, while being largely unchanged on further warming into $1T^{\prime}$ \cite{tao_appearance_2019}, which mimics the trends in $K_x$.)
The possibility of the band structure determining the interlayer vibrational coupling has implications for the stacking energetics.
The free energy is a function of the vibrational and electronic band structure. However, if the interlayer vibrational coupling can be modified by $\sim$20\% by stacking changes, then the effect of stacking-dependent changes in the band structure may need to be carefully considered (i.e., with calculations precise enough to compute realistic values of $\hbar \omega_m$) before the vibrational contribution to the free energy can be properly evaluated.

Of course, with our use of the linear chain model, we have made some assumptions that should be investigated further. First, are the layers essentially identical, or are deviations in intralayer atomic positions between the layers important for the stacking energetics or other properties? 
Are the intralayer vibrations that may complement the LCM modes the same in every layer?
Second, our results may hold in the bulk, but how do the properties of surface layers and thin films of Mo$_{1-x}$W$_{x}$Te$_{2}$ differ from bulk samples? It is known that the transition of MoTe$_{2}$ is broadened or suppressed for thin samples \cite{cao_barkhausen_2018,he_dimensionality-driven_2018,paul_tailoring_2020}. There is some evidence for weaker interlayer vibrational coupling for few-layer samples; the interlayer force constants $K_x$ from Raman measurements on $
\leq$8-layer MoTe$_{2}$ are 0.673(11) and 0.604(15) for $T_d$- and $1T^{\prime}$-MoTe$_{2}$, respectively  \cite{cheon_structural_2021}, both substantially smaller than our values of $0.919(24)$ and 0.760(15) for bulk Mo$_{0.91}$W$_{0.09}$Te$_{2}$. Also, bilayer WTe$_{2}$ shows signs of a transition above $\sim$340 K (in the disappearance of a second harmonic generation signal \cite{fei_ferroelectric_2018}); if the intralayer positions are unchanged, then the only explanation for the arrival of inversion symmetry in a bilayer structure would be a structure with $\delta=0.5$ (analogous to the hypothetical $T_0$ phase discussed in Ref.\ \cite{huang_polar_2019}), which would require a substantial change of interlayer vibrational coupling compared to bulk samples. (A transition from $T_d$ to $1T^{\prime}$ in bulk WTe$_{2}$ has been observed near $\sim$560 and 613 K \cite{tao_t_d_2020,dahal_tunable_2020}, but the $\delta$ parameter is largely unchanged across this transition \cite{tao_t_d_2020}.)
Of course, the tendency for the transition to be suppressed due to insufficient thickness, and the gradual suppression of stacking-related diffuse scattering on either warming into $1T^{\prime}$ or cooling into $T_d$ \cite{tao_appearance_2019}, further indicates the importance of long-range interlayer interactions to the stacking energetics. 

Stacking energetics are of prime importance for many quasi-2D materials, but they are still poorly understood. 
Ideally, we could obtain insight from studies on graphite, which is another layered semimetal that can have multiple stacking variations, and where the relative positioning of neighboring layers is the same regardless of overall stacking. 
It is curious how Bernal-stacked graphite is dominant in nature, despite the weakness of the interlayer interactions.
However, despite the attention that graphite/graphene has received and the simplicity of its structure, the energy differences between different stacking possibilities in graphite are not well understood. For example, DFT calculations have been inconsistent on whether Bernal or rhombohedral graphite has the lower free energy at room temperature \cite{charlier_first-principles_1994,anees_ab_2014, savini_bending_2011,taut_electronic_2013,nery_ab-initio_2021}. 
There has been some focus on how changes in the electronic structure affect the free energy, with electronic temperature argued to be essential to determining which graphite stacking is preferred at a certain temperature \cite{nery_ab-initio_2021}. Meanwhile, the vibrational contribution to the stacking-dependence of the free energy in graphite tends to be neglected. There is some evidence that the interlayer modes of trilayer graphene are $\sim$1-2\% softer for rhombohedral-like than Bernal-like stacking \cite{lui_stacking-dependent_2015}, so it would be interesting to see how changes in the vibrational spectra with stacking effect the free energy in graphite.
Indeed, our results show that the interlayer vibrational coupling of a van der Waals layered material can change substantially between phases of different stacking. 

The possible connection between the band structure and interlayer vibrational coupling may yield insight into how the transitions in Mo$_{1-x}$W$_{x}$Te$_{2}$ are effected by optical or electronic means; such means include pulses of light in ultrafast spectroscopy \cite{sie_ultrafast_2019}, an electron beam \cite{huang_polar_2019}, and an applied electric field (for few-layer WTe$_{2}$)  \cite{fei_ferroelectric_2018, xiao_berry_2020}. Additionally, there are other materials that exhibit stacking transitions in few-layer films even when not seen in the bulk; such transitions can be induced with an applied electric field on bilayer hexagonal boron nitride \cite{yasuda_stacking-engineered_2021}, and with laser irradiation on trilayer graphene \cite{zhang_light-induced_2020}. 
Given the difficulty of calculating properties that depend on the weak interlayer interactions, our finding that the interlayer vibrational coupling can change by $\sim$20\% between differently-stacked phases should provide insight into how stacking transitions may occur in a wide range of other systems.

It seems unusual that there is such a large change in an elastic constant (namely, $C_{55} = K_x t$, where $t$ is the interlayer spacing \cite{grzeszczyk_raman_2016}) between phases. Comparable changes have been seen in NiTi in the vicinity of its martensitic transition \cite{grabec_evolution_2021}, and graphite does have a greatly increased $C_{55}$ constant after irradiation \cite{ayasse_softening_1979}, but Mo$_{1-x}$W$_{x}$Te$_{2}$ may be unique in being a van der Waals layered system with reversible changes in $C_{55}$ of the magnitude observed. Furthermore, since a sufficiently strong applied electric field can induce stacking changes in few-layer WTe$_{2}$ \cite{fei_ferroelectric_2018,xiao_berry_2020}, it may be worth investigating if a smaller electric field can modulate the interlayer vibrational coupling, which would open an avenue of research into whether elastic properties can be modulated by electrical means. 
Additionally, if changes in the band structure are responsible for the changes in the elastic constant $C_{55}$, then it may, conversely, be possible to modify the Weyl dispersion by applying a shear stress to Mo$_{1-x}$W$_{x}$Te$_{2}$. 
Our results suggest a coupling between the elastic/vibrational properties and the interlayer electronic structure which should prove a fruitful avenue for future exploration.

\section{Conclusion}

We performed inelastic neutron scattering measurements to observe the $a$-axis interlayer shear mode phonons in the $T_d$, $T_d^*$, and $1T^{\prime}$ phases. The phonon peak positions were consistent with the linear chain model, though there is a substantial difference in the interlayer force constants between the phases, with the $K_x$ values of $T_d^*$ and $1T^{\prime}$ about 76(3)\% and 83(3)\% that of $T_d$. 
The large change in $K_x$, in contrast to the small changes in the $\delta$ (or $\epsilon$) parameters or the intralayer positions, suggests that stacking-induced changes in the electronic band structure may be responsible for the change in vibrational properties.

\section*{Acknowledgements}

This work has been supported by the Department of Energy, Grant number
DE-FG02-01ER45927.  A portion of this research used resources at the High Flux Isotope Reactor and the Spallation Neutron Source, which are DOE Office of Science User Facilities operated by Oak Ridge National Laboratory. We acknowledge the support of the National Institute of Standards and Technology, US Department of Commerce, in providing neutron research facilities used in this work.

\nocite{fobes_neutronpy_2020,zheludev_reslib_2007-2,cooper_resolution_1967,rano_ab_2020,singh_engineering_2020,tao_appearance_2019}


%

\end{document}